\documentclass[10pt,pre,twocolumn,superscriptaddress,showpacs]{revtex4-1}
\usepackage[utf8]{inputenc}
\usepackage{amsmath,amssymb}
\usepackage{amsfonts}
\usepackage{graphicx}
\usepackage{color}
\usepackage[dvipsnames]{xcolor}

\newcommand{\mb}{\mathbf}

\begin{document}

\title{Unsteady flow, clusters and bands in a model shear-thickening fluid}
\author{Shibu Saw} \affiliation{Institute for Theoretical Physics,
  Georg-August University of G\"ottingen, Friedrich-Hund Platz 1,
  37077 G\"ottingen, Germany} 
\author{Mathias Grob} \affiliation{Institute for Theoretical Physics,
  Georg-August University of G\"ottingen, Friedrich-Hund Platz 1,
  37077 G\"ottingen, Germany}
\author{Annette Zippelius} \affiliation{Institute for Theoretical Physics,
  Georg-August University of G\"ottingen, Friedrich-Hund Platz 1,
  37077 G\"ottingen, Germany}
\author{Claus Heussinger}\affiliation{Institute for Theoretical Physics, Georg-August
  University of G\"ottingen, Friedrich-Hund Platz 1, 37077
  G\"ottingen, Germany}

\begin{abstract}
  We analyse the flow curves of a two-dimensional assembly of granular
  particles which are interacting via frictional contact forces.  For
  packing fractions slightly below jamming, the fluid undergoes a
  large scale instability, implying a range of stress and strainrates
  where no stationary flow can exist. Whereas small systems were shown
  previously to exhibit hysteretic jumps between the low and high
  stress branches, large systems exhibit continuous shear thickening
  arising from averaging unsteady, spatially heterogeneous flows. The
  observed large scale patterns as well as their dynamics are found to
  depend on strainrate: At the lower end of the unstable region, force
  chains merge to form giant bands that span the system in
  compressional direction and propagate in dilational direction.  At
  the upper end, we observe large scale clusters which extend along
  the dilational direction and propagate along the compressional
  direction. Both patterns, bands and clusters, come in with infinite
  correlation length similar to the sudden onset of system-spanning
  plugs in impact experiments.

\end{abstract}

\maketitle
\section{Introduction}

The defining feature of non-Newtonian fluids is that the viscosity,
i.e. the resistance of the fluid to flow, is not a material constant
but depends on the flow itself. A shear thickening fluid, in
particular, has the property that the viscosity increases with the
speed of the flow. Shear thickening (ST) can be modest, with only a
small increase of the viscosity above its zero-strainrate value. It
may also be an order of magnitude effect, and even lead to a
discontinuous flow arrest upon increasing the driving force beyond a
threshold. Such a spectacular phenomenon is quite opposite to
``normal'' materials that start to flow or break upon increasing the
force.

A related phenomenon is the establishment of system-spanning solid
structures upon impact of an object on the
surface~\cite{maharjan18:_const,allen18:_system}. These structures may
even be strong enough to carry persons, at the same time absorbing
enough kinetic energy to be used as
shock-absorbers~\cite{decker07:_stab_stf}.

It has been shown by
experiments~\cite{brown12JRheol,0034-4885-77-4-046602, Clavaud5147,comtet,PhysRevE.92.032202,Hsu5117}
and simulations \cite{ClausPRE2013,PhysRevLett.111.218301,C5SM02326B}
that solid-solid friction between particles is the relevant force for
the ST effect. The onset for ST is governed by an intrinsic force
scale, sometimes modeled via a switch, at which frictional forces
start to be
relevant~\cite{PhysRevLett.111.218301,PhysRevLett.112.098302}. The
upper limit for ST is equally set by a force-scale, that represents
the weakest link in the system, e.g. surface tension between the
sample and surrounding air in a
rheometer~\cite{0034-4885-77-4-046602}. In simulations this force can
be fixed by the stiffness of the particles, which leads to a
finite-yield stress even in the arrested state (in contrast to
hard-sphere systems)~\cite{ClausPRE2013}. The resulting phase diagram
has a re-entrant shape with a fluid-solid transition at low stress and
a re-fluidization at higher stress~\cite{GrobPRE2014}. Several
variations of the basic system have been considered, including
e.g. inertia~\cite{PhysRevLett.111.108301,GrobPRE2014} or Brownian
forces~\cite{Mari15326}. Subsequently, it turned out that ST
represents an unsteady coexistence of different fluid
states~\cite{Grob2014,0295-5075-115-5-54006}. Vorticity banding has
also been observed~\cite{PhysRevLett.121.108003}, albeit with dynamic
bands that move along the vorticity direction. Experiments also
observe unsteady states, e.g. Saint-Michel {\it et  
  al.}~\cite{PhysRevX.8.031006} report propagating bands at the onset
of ST, and a proliferation of the dynamics deeper in the ST
state. Rathee {\it et al.}~\cite{rathee17:_local} also find localized
regions of increased stress that occur intermittently. Interestingly
the size of these regions seem to grow with the gap-width of the
rheometer, i.e. with the system size.

In this study we characterize the unsteady state of a model ST
fluid. We build on our previous work published in
Refs.~\cite{Grob2014,GrobPRE2014}.

\section{Model}\label{sec:model}

Our starting point are Newton's equations of motion for
a 
granular mixture of 
dry frictional 
particles in two space dimensions. Forces between particles arise at contact and are frictional. The particles are modeled as soft spheres of radius $R_i$, interacting with normal and tangential forces: 
\begin{eqnarray}
{\bf f}_{ij} = {\bf f}_{ij}^{(n)} + {\bf f}_{ij}^{(t)}.  \label{eq-f}
\end{eqnarray}
Denoting particles' positions and velocities by $\{{\bf r}_i\}$ and $\{{\bf v}_i\}$, the visco-elastic
normal force can be written as~\cite{silbert}
\begin{eqnarray}
  {\bf f}_{ij}^{(n)} &=& \Big(k^{(n)} \delta_{ij}^{(n)} - \eta^{(n)}v_{ij}^{(n)}\Big) \Theta(\delta_{ij}^{(n)}-r_{ij})  {\bf n}_{ij}.   \label{eq-fn}
\end{eqnarray}
The unit vector, ${\bf n}_{ij} \equiv {\bf r}_{ij}/r_{ij}$, points from the center of particle $i$ to the center of particle $j$ and the particles only interact, when they overlap, $\delta_{ij}^{(n)} \equiv R_i+R_j-r_{ij}>0$.
The $k^{(n)}$ and $\eta^{(n)}$ are the elastic and damping
coefficients along  the normal direction and 
$v_{ij}^n \equiv ({\bf v}_i-{\bf v}_j)\cdot{\bf n}_{ij}$ denotes the normal component of the relative velocity.



The tangential force between particle $i$ and $j$ is given
by~\cite{silbert}
\begin{eqnarray} {\bf f}_{ij}^{(t)} &=& \mathrm{min} \Big(|k^{(t)}
  \delta^{(t)} - \eta^{(t)} v_{ij}^{(t)}|, \mu |{\bf f}_{ij}^{(n)}|
  \Big) \;{\bf t}_{ij}  \nonumber\label{eq-ft}
\end{eqnarray}
where $k^{(t)}$ and $\eta^{(t)}$ are the elastic and damping
coefficients along tangential direction ${\bf t}_{ij}$, defined by
${\bf t}_{ij}\cdot{\bf n}_{ij}=0$. The function $\mathrm{min(m,n)}$
yields the lower value between $m$ and $n$, enforcing the Coulomb
criterion $|{\bf f}_{ij}^{(t)}| \le \mu |{\bf f}_{ij}^{(n)}|$ with the
friction coefficient $\mu$. The relative tangential velocity at
contact
$v_{ij}^{(t)} = ({\bf v}_i - {\bf v}_j)\cdot{\bf t}_{ij} +
(R_i\omega_i + R_j\omega_j)$ is the sum of a translational and a
rotational contribution.  It determines the tangential displacement
according to
$\delta_{ij}^{(t)} = \int_{t}^{t+\Delta t} dt ~ v_{ij}^{(t)}$, where
the integration is over the time ineterval $[t,t+\Delta t]$, when the
contact is present and not sliding.


We consider an equi-quaternary mixture with particle sizes:
$2R_{A}=0.7$, $2R_{B}=0.8$, $2R_{C}=0.9$ and $2R_{D}=1.0$, all of equal mass $m$.
The units of length, time and stress are chosen as $2R_D$, $(m/k^{(n)})^{1/2}$ and $k^{(n)}$.
The friction coefficient has been set to
$\mu=2$ in accordance with previous work~\cite{Hayakawa,Grob2014} and the damping constants are set to
$\eta^{(n)}=\eta^{(t)}=1/2$.

A constant strainrate $\dot\gamma $ along the $x$-axis is imposed
with help of Lee-Edwards boundary condition.
 Molecular dynamics simulations have been performed using the LAMMPS simulation
package~\cite{lammps,silbert} for
$N=8000$ up to $N=80000$ particles at various
packing fractions $\phi$.

\section{Results and Discussion}\label{sec:results}

\subsection{Global stress-strain relation}
\begin{figure}[t]
	\includegraphics[width=0.4\textwidth]{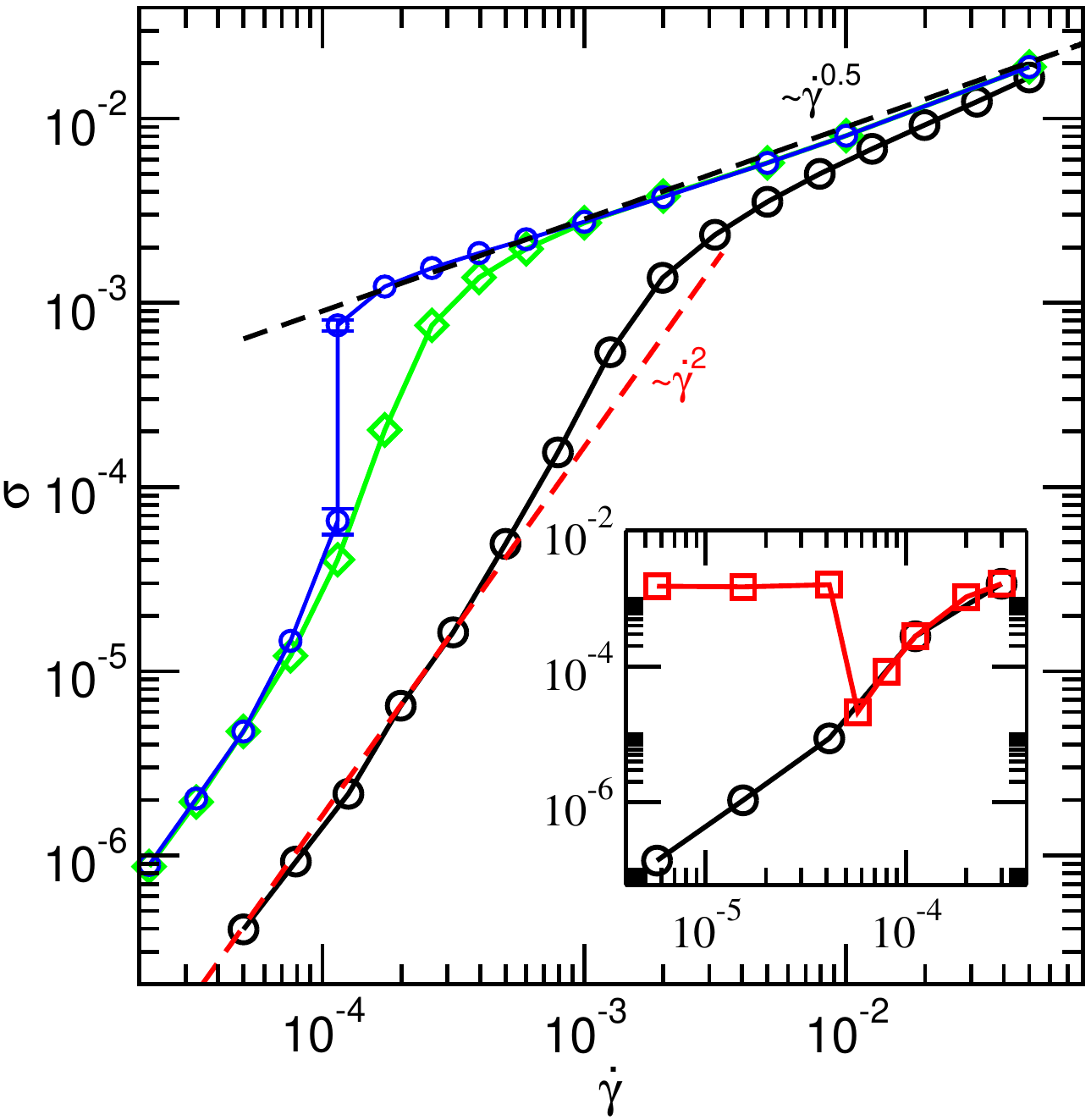}
        \caption{Flow curves for different system sizes $N$ and volume
          fractions $\phi$: ($\phi=0.78,N=3600$, black),
          ($\phi=0.7975,N=8000$, blue, and $N=32000$, green), (inset
          $\phi=0.801,N=80000$).
          Inset: different starting configurations; the yield-stress
          branch is metastable.}
  \label{fig:flowcurve}
\end{figure}

We want to understand the heterogeneous, time-dependent shear stress,
which develops in response to an applied strainrate in large
systems. Before analysing these structures in detail, we briefly
recall the global stress-strain relations and discuss the non-trivial
dependence on system size. Some representative flow curves are shown
in Fig.~\ref{fig:flowcurve}. One observes the well-known Bagnold
scaling at small $\dot\gamma $: $\sigma=\eta \dot\gamma^2$ and
Herschel-Bulkley (HB) like behaviour $\sigma\propto \gamma^{1/2}$ for large
$\dot\gamma$. For ``small'' volume fraction (black curve), these two
regimes are connected by a smooth crossover and ST is relatively
weak. For increasing volume fraction $\phi$, one observes stronger
shear thickening, which is discontinuous for the small system (blue
curve) and continuous for the large one (green curve). For even larger
$\phi$ a finite yield stress is required for the system to
flow (red curve, inset).

\begin{figure}[t]
	\includegraphics[width=0.4\textwidth]{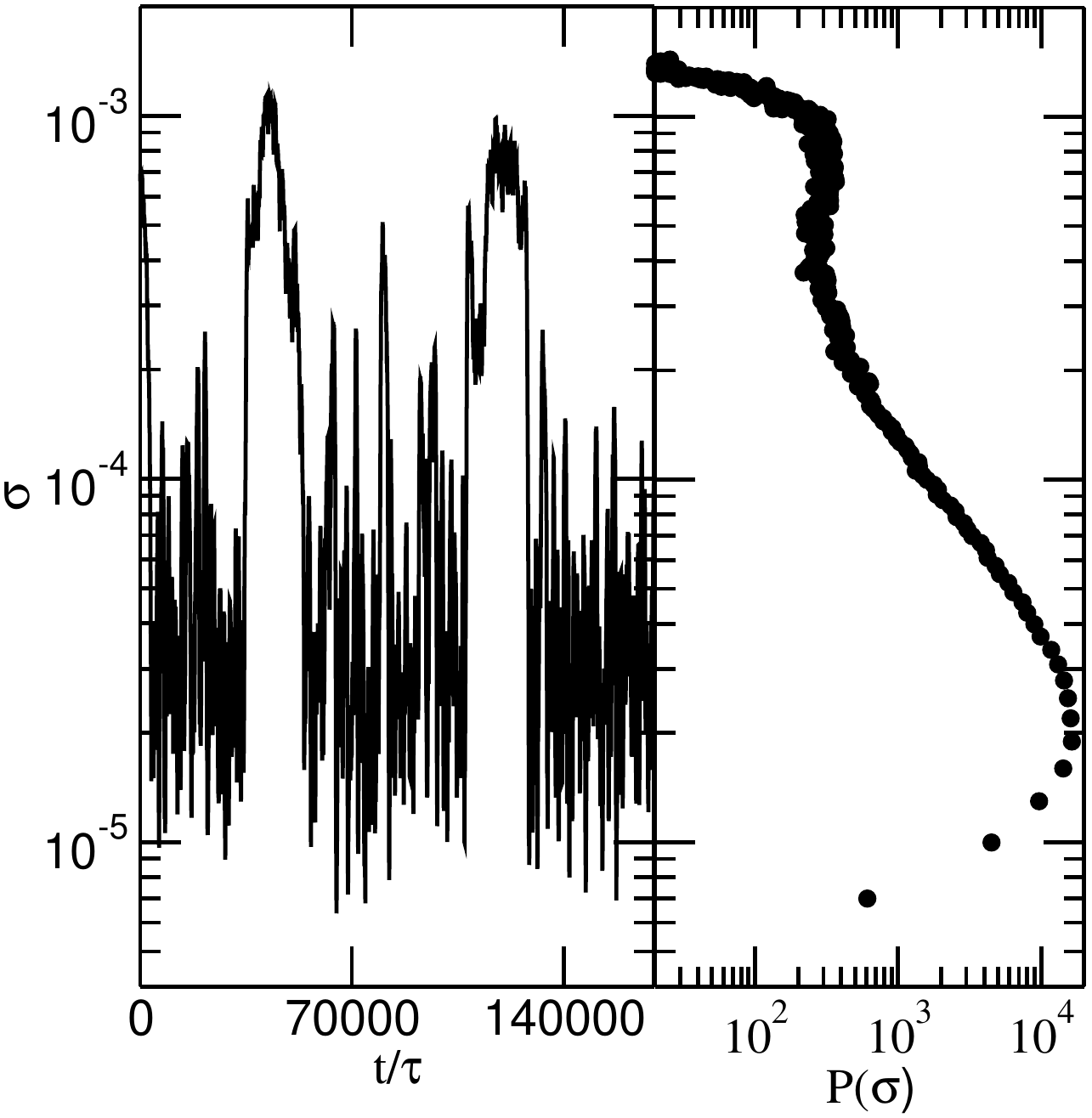}
        \caption{Left: stress $\sigma$ as a function of time for
          $N=8000$, $\phi=0.7975$ and
          $\dot\gamma=1.143\times 10^{-4}$; right: corresponding
          distribution of stress, $P(\sigma)$. The average over this
          distribution, restricted to the regions around the two peaks
          gives the two data points in the blue curve of
          Fig.~\ref{fig:flowcurve}, that carry an error bar. The error
          is obtained by varying the stress value at which the
          distribution is split.}
        \label{fig:jumpstatistics}
\end{figure}

Discontinuous ST is visible as a sudden jump of the stress from the
fluid to the HB branch, for $\phi=0.7975$ at a strainrate
$\dot\gamma_l\approx 10^{-4}$. There is clearly a range of forbidden
values of $\sigma$ and when observed with temporal resolution, one
finds that the small system jumps frequently between the two
branches. In Fig.~\ref{fig:jumpstatistics} we show such a trajectory
together with the distribution of $\sigma$-values for this
run. Clearly the distribution is bimodal with the two peaks
corresponding to the low $\sigma$ Bagnold regime and the high $\sigma$
HB regime. The jump is associated with hysteresis within a finite
range of strainrates, $\dot\gamma_l\leq\dot\gamma \leq \dot\gamma_u$,
when comparing a ramping simulation with slowly increasing strainrate
(jump at $\dot\gamma_u$) with simulations with decreasing strainrate
(jump at $\dot\gamma_l$).

These features bear some resemblance with the phenomenon of phase
coexistence, where the jump represents the forbidden (unstable) region
in the $\sigma-\dot\gamma$ phase diagram, and the ``critical point'' lies
in the flowcurve with diverging slope
$d\sigma/d\dot\gamma\to\infty$. However, as it turns out there is a
non-trivial dependence on system-size which is not expected for that
type of phenomena.  As can be seen in Fig.~\ref{fig:flowcurve}, the
discontinuity in the flow-curve is a finite-size effect and vanishes
if the system is large enough. In thermodynamic systems with phase
coexistence on the other hand, larger systems imply longer time scales
for the nucleation of domain walls, and therefore a more pronounced
discontinuity or hysteresis in finite-time simulations.

What happens in large systems? There is still a range of forbidden
$\sigma$-values in the sense that in this range no stationary
homogenous flow exists. Instead we observe heterogeneous
time-dependent flow, whose properties will be analysed in the
following paragraphs. The smooth curve in Fig.~\ref{fig:flowcurve}
does not represent stationary flow, but is obtained by averaging the
flow over space and time.

The range of forbidden $\sigma$-values, depends on volume fraction and
so does the the critical system-size for the vanishing of the
discontinuity. While for $\phi=0.78$ the flowcurve is continuous
already for $N=3600$, a much larger system ($N=32000$) is needed for
the volume fraction $\phi=0.7975$. At even higher $\phi=0.801$ (inset)
a discontinuity persists up to $N=80000$.

Interestingly, at this volume fraction one observes a discontinuity
that does not lead upwards in stress when increasing the strainrate,
but downwards (red curve, inset). Here, a metastable yield-stress
branch becomes unstable at $\dot\gamma\approx 5\cdot 10^{-5}$. After
the instability, at higher strainrates, the system follows a flowcurve
that resembles the heterogeneous time-dependent flows that are
displayed in the main panel. This downward jump is remarkable for two
reasons. First, its presence depends on initial conditions. Choosing
different starting configurations, one also observes a continuous
branch (black curve, inset) which, instead of a yield-stress, shows
Bagnold behavior at small strainrates. We expect the time-scale that
governs this memory to diverge with increasing system size and
also when crossing the jamming transition. Secondly, a downward jump
opens room for additional instabilities in the form of
shear-bands. Such a phenomenology of ``discontinuous shear-thinning''
has recently been discussed by one of us in Ref.~\cite{2018irani}.

As a result of this section we conclude that ST in infinite systems
always seems to be continuous, but flow is spatially heterogeneous and
time-dependent. The actual system size that is necessary to reach this
limit depends on volume fraction. For smaller volume fractions, like
for the black curve in Fig.~\ref{fig:flowcurve}, the relatively small
system ($N=8000$) is already large enough. Thus, the unsteadyness
might already be present at the lowest volume-fractions that show only
mild ST and could therefore be an inherent feature of any ST flow.

\subsection{Heterogenous stress states}

\begin{figure}
  \includegraphics[width=0.42\textwidth]{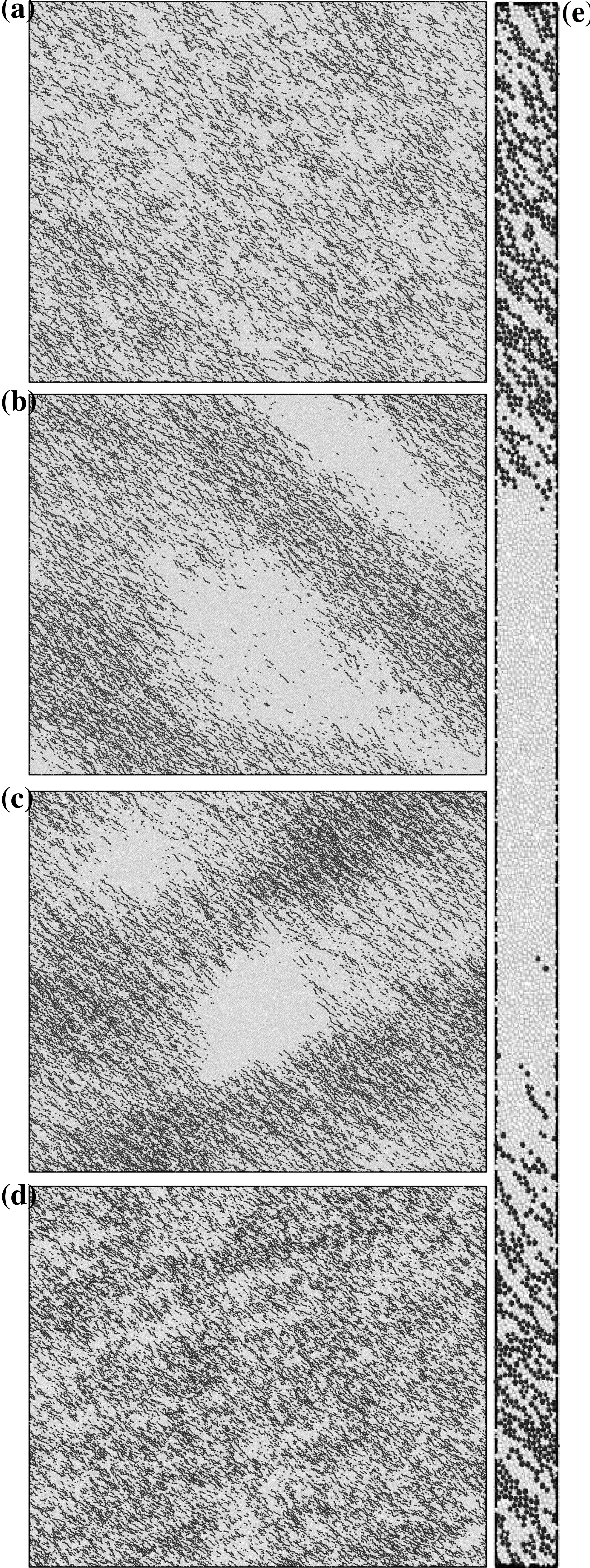}
  \caption{Snapshots of local shear stress for (a)
    $\dot\gamma=3.297\times10^{-5}$, (b)
    $\dot\gamma=1.143\times10^{-4}$, (c) $\dot\gamma=10^{-3}$ and (d)
    $\dot\gamma=0.05$. Particle $i$ is colored in black (gray),
    whenever its local shear stress $\sigma_i$ is above (below)
    average; (e) snapshot of local shear stress for a high aspect
    ratio sample: $L_x=10, L_y=260$
    ($\dot\gamma=1.732\times10^{-4}, \phi=0.79875$) }
\label{local_shear}
\end{figure}

To better understand the heterogenous stress states in large systems,
we analyse representative snapshots, such as those displayed in
Fig.~\ref{local_shear}. The local shear stress of each particle is
calculated as
$\sigma_i=\sum_jF_{ij}^xy_{ij}+v_i^y(v_i^x-y_i\dot\gamma)$, where
$F_{ij}^x$ is the x-component of the interaction force between
particles i and j, and $y_{ij}=y_i-y_j$ the distance in
y-direction. Except for the largest strainrates, the kinetic
contribution to the stress is negligible. Particles are colored black
if their local stress exceeds the average value, and gray otherwise.

In the flowing state (panel a) linear structures that bear high
stresses can be seen. These forcechains are oriented along the
compressive direction of the flow (flow in positive x-axis) and seem
to have a typical size of 5-10 particles. This length presumably
depends on volume fraction and diverges at
jamming~\cite{PhysRevE.76.021302}. The forcechains are distributed
homogeneously throughout the system and seem to exist independently
from each other. Similarly, for the highest strainrates (panel d),
the flow is approximately homogeneous and time independent

In the continuous shear thickening (CST) region, on the other hand,
the distribution of forcechains is rather inhomogeneous (b and c) and
we observe large patches of high-stresses coexisting with regions of
small stresses. Thus, the emerging CST in large systems is reminiscent
of spatial coexistence of an inertial flow state (small-stress) and a
plastic flow state (large-stress). However, coexistence cannot
represent stationary two-dimensional flow. Hence the observed patterns
are inherently time-dependent, as will be discussed in the next
section.  In small systems, spatial coexistence is not
possible. Instead the whole system corresponds to a patch of either
inertial or plastic flow and switches as a function of time between
these two homogeneous states.

If we look in more detail then we can distinguish different
features. Panel b is taken at a strainrate
$\dot\gamma_l\approx 10^{-4}$ at the lower end of the CST
region. Here, the force-chains seem to merge together to form giant
bands that span the system in compressional direction, i.e. parallel
to the chains themselves. Panel c on the other hand corresponds to a strainrate
$\dot\gamma_u\approx 10^{-3}$ which is one order of magnitude larger than
$\dot\gamma_l$ and marks the upper end of CST. Here, the structures do
not seem to percolate in the compression direction but rather form
clusters that extend along the dilational direction. Thus, the type of
structures that form (clusters, bands) depends on and changes with the
strainrate~\cite{doi:10.1063/1.166456}.

To put these observations on a quantitative basis, we calculate the
stress-stress spatial correlation function
$C_{\sigma}({\mb r})=\frac{1}{N}\sum_{i\neq j}\langle \sigma_i\sigma_j
\delta({\mb r}-{\mb r}_{ij})\rangle$.
It is obvious from Fig.~\ref{local_shear} that the correlations in the
CST regime are anistropic as has been observed previously in
experiment~\cite{majmudar05:_contac}. To capture these anisotropies we
consider correlations in the compressive and dilational directions
separately by restricting the separation vector ${\mb r}$ to the
compressive and dilational direction, respectively. A length-scale
$\xi$ can be extracted by monitoring the distance, at which the
correlation function drops to a certain fraction
($C(\xi)=0.07\cdot C(0)$)). This value needs to be chosen small
enough to not interfere with short distance effects, and large enough
to avoid noise and finite-size effects due to the large distances
involved. Within these bounds the results presented are independent of
the value chosen.

\begin{figure}[h!]
  \includegraphics[width=0.23\textwidth]{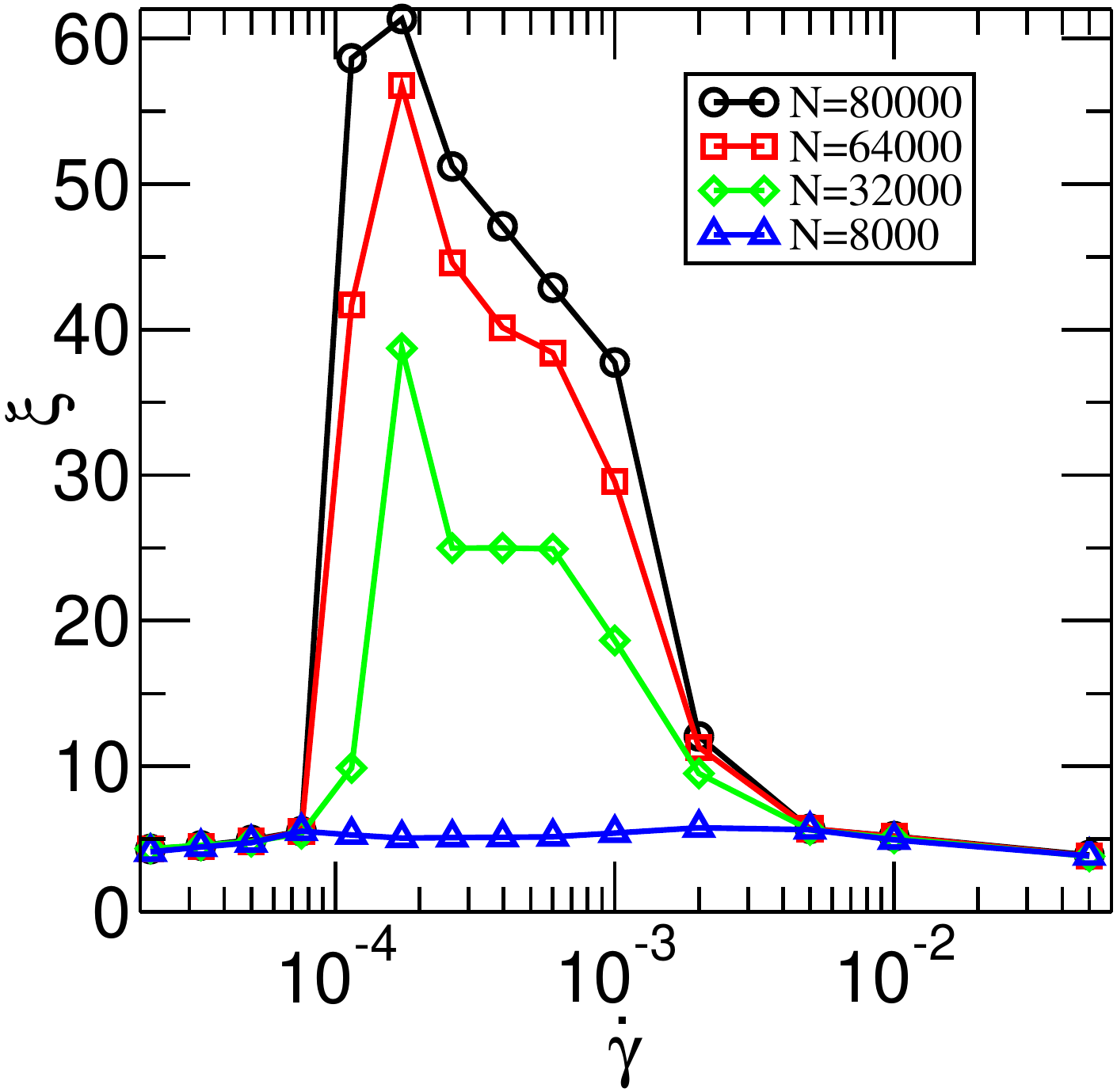}
  \hfill
  \includegraphics[width=0.23\textwidth]{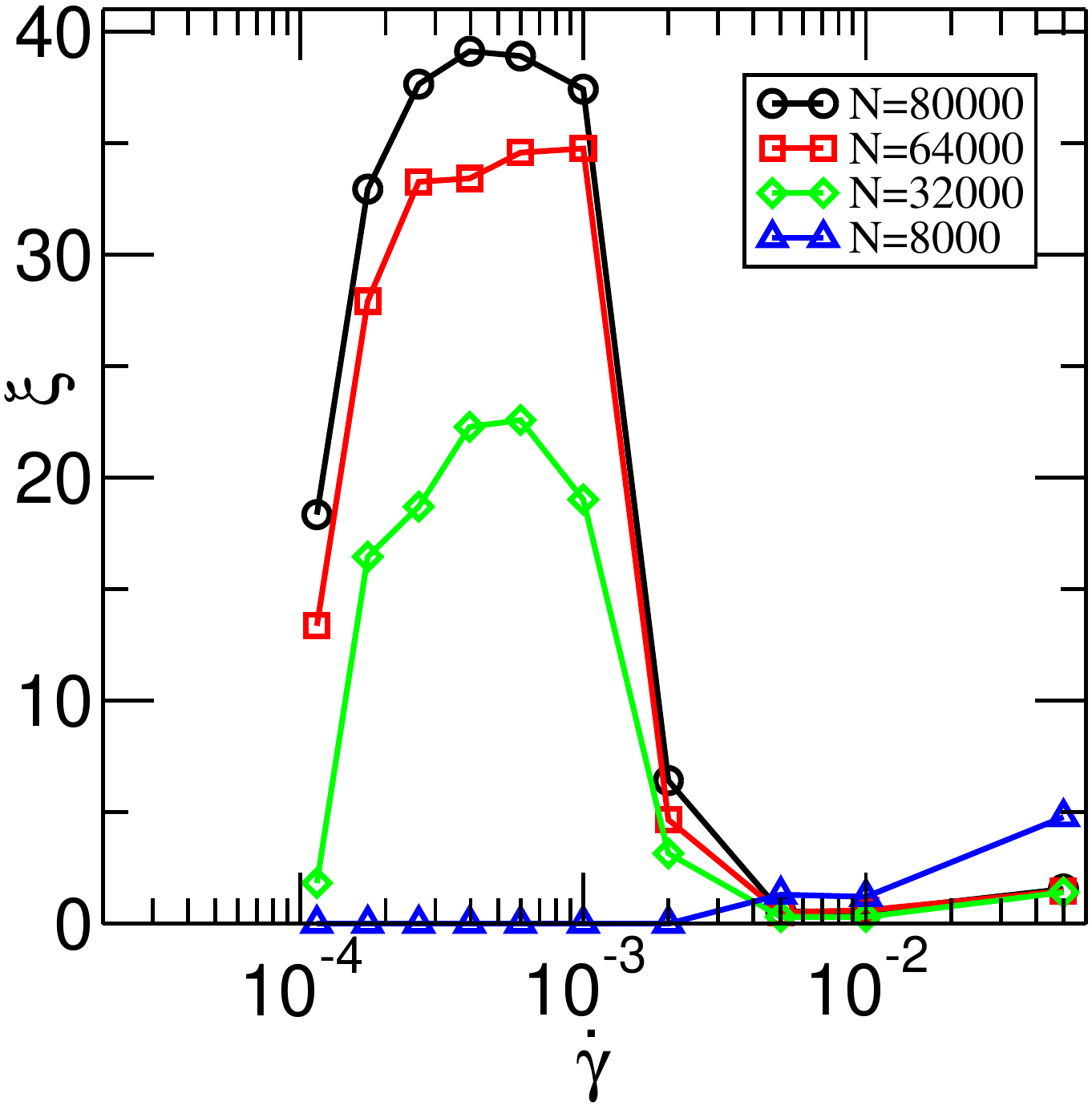}
  \caption{Correlation length $\xi$ vs strainrate
    $\dot\gamma$. Length-scale extracted from the decay of the stress
    correlation function $C_\sigma(r)$ with $r$ taken along the
    compressive (left) or the tensile (right) direction.}
  \label{fig:stress_diagonal_0.79750}
\end{figure}

The resulting correlation lengths in these directions are plotted in
Fig.~\ref{fig:stress_diagonal_0.79750}. Different system sizes are
included. For the smallest system $N=8000$ no length scale is observed
in the range of intermediate strainrates, where the CST regime
resides. With an eye on Fig.~\ref{fig:flowcurve} we confirm that this
system does not show CST, but rather the discontinuous jump from
Bagnold to HB. Once the system is large enough to show CST a large length-scale
builds up that ever increases with system size. In fact, we find
$\xi\propto L$, with $L$ the linear dimension of the simulation box. 

The peak value of the correlation length is different in the two
directions. In the compressive direction the maximum length is
obtained at the lower end of CST, at $\dot\gamma_l$. In dilational
direction, maximal correlation is achieved deep within CST and towards
$\dot\gamma_u$.  This proofs the visual impression from
Fig.~\ref{local_shear} that structures change their orientation.

To understand these local correlations in more detail, we also looked
at other fields, such as the local pressure, the local density and the
local connectivity. In Fig.~\ref{fig:stress_diagonal_0.79750} we show
a profile of the local shear stress and the local connectivity along
the compressive direction, averaged over the dilational direction.
Stress $\sigma$ and particle connectivity $z$ are seen to be strongly
correlated, as one might expect, because stresses are carried by
contacts between the particles. The figure suggests a relation
$\sigma=\sigma_0\exp(z/z_0)$ with constants
$\sigma_0\approx 5\cdot10^{-7}$ and $z_0\approx0.4$, thus stress is
exponentially amplified if local connectivity is large enough.

Correlations of stress and density are also present, but less
pronounced. A quantitative calculation gives a positive correlation
coefficient of 0.07 for stress and density, when coarse-graining the
fields on a grid with size 2. However, a small correlation might
already induce strong effects. At volume-fractions close to the
singular jamming point only small variations in density are necessary
to change the character of the flow substantially. This is sometimes
used as argument to favor pressure-controlled over volume-controlled
flows, as such a singularity is not present when pressure is
controlled~\cite{guazzelli_pouliquen_2018}.

\begin{figure}[h]
  \includegraphics[width=0.4\textwidth]{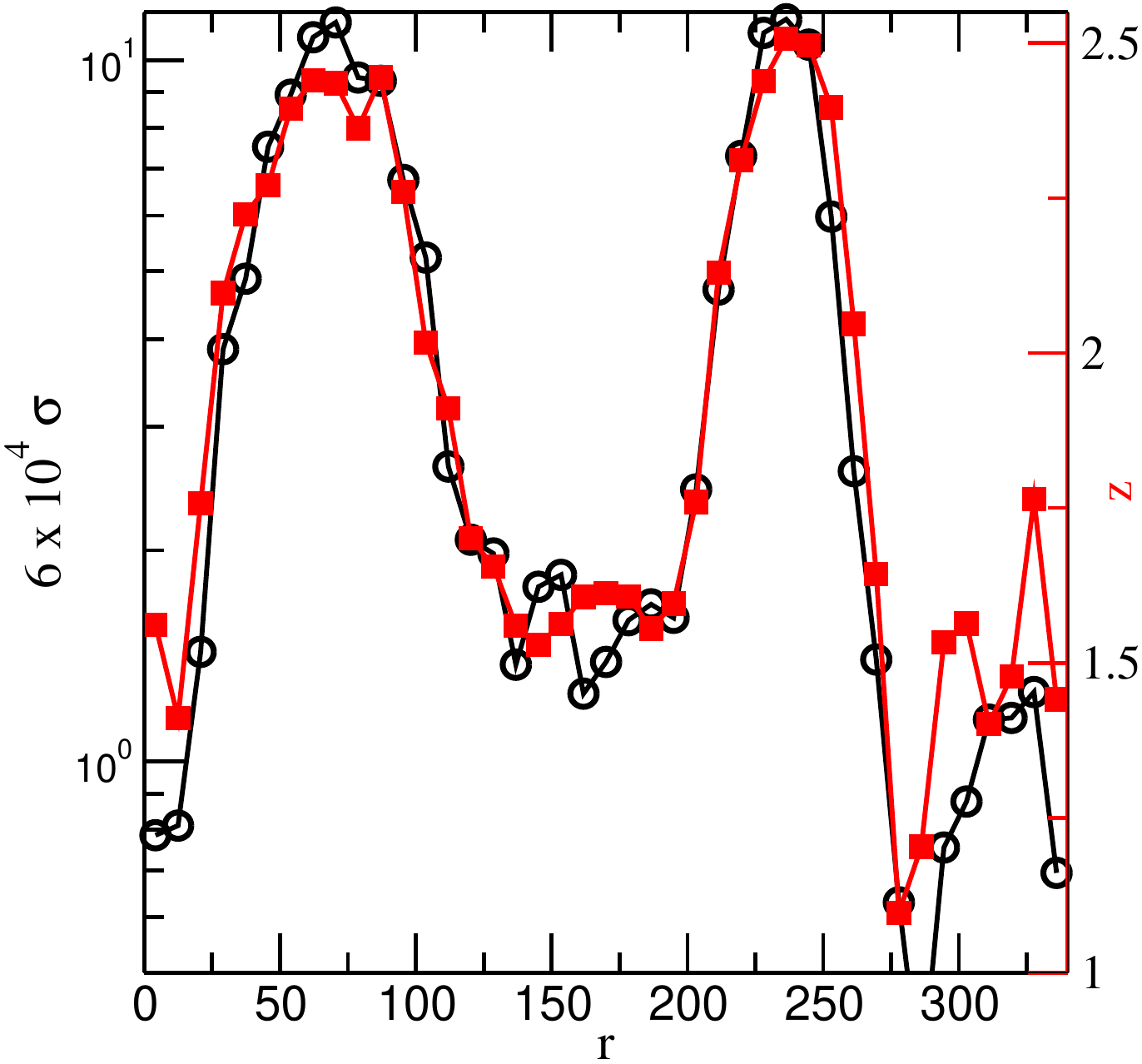}\vspace{2mm}
  \caption{Profiles of local shear stress $\sigma$ and number of
    contacts per particle $z$, taken along the (compressive) diagonal
    direction $r$. Values are averaged over tensile direction.}
  \label{fig.localprop}
\end{figure}

We have also investigated different aspect ratios $L_x/L_y$ of the
simulation box. Remarkably CST can also be observed for systems with
only few particles, provided the system is large in the gradient
direction ($L_y\gg L_x$). In the right part of Fig.~\ref{local_shear}e
we show the local shear stress in a system of only 3000 particles with
$L_x=10$ and $L_y=216$. The temporary shear banding is clearly seen
and the global stress follows the same continuous flow curve (CST), as
displayed in Fig.\ref{fig:flowcurve} for the large system (green
curve). At the same, a system with an equivalent number of particles
but in a square simulation box ($N_x=N_y$) would feature a
discontinuous flow-curve (as with the blue curve).

To conclude this section, CST is due to a large-scale coexistence of
high- and low-stress regions. Most remarkable is the seemingly
discontinuous onset of the system-spanning bands in compressive
direction at the lower end of CST. This discontinuity exists whenever
systems are large enough in gradient direction. Still the flowcurves
are continuous. It is tempting to relate this to the system-spanning
plugs that occur in impact experiments~\cite{allen18:_system}. These
plugs only occur above an onset stress and represent a solidified
state of the shear-thickening suspension.


\subsection{Dynamics of shear bands}

The structures visible in Fig.~\ref{local_shear} are not static but
dynamically evolving. Interestingly, we can observe propagating modes
with structures moving with a certain velocity. The clusters in panel
c move in the direction parallel to the forcechains they are made of,
i.e. in compressive direction. They do not stay coherent over a long
time, rather they constantly split and join together over the course
of time.

The system-spanning bands visible in panel b move perpendicular to
their main axis, i.e. in dilational direction. Bands sometimes seem to
evolve out of clusters that join together, percolating the entire
system in compressive direction. Motion in this direction then stops
and motion in dilational direction sets in. During propagation, the
bands do not rotate but keep their orientation. This implies that the
Lees-Edwards periodic boundary conditions in y-direction impose some
offset that tends to destabilize the bands. Indeed, some time after
its establishment we can observe deformations of the band until it
splits in two or more parts. These parts then behave similar to the
clusters described above.

\begin{figure}[h]
	\includegraphics[width=0.4\textwidth]{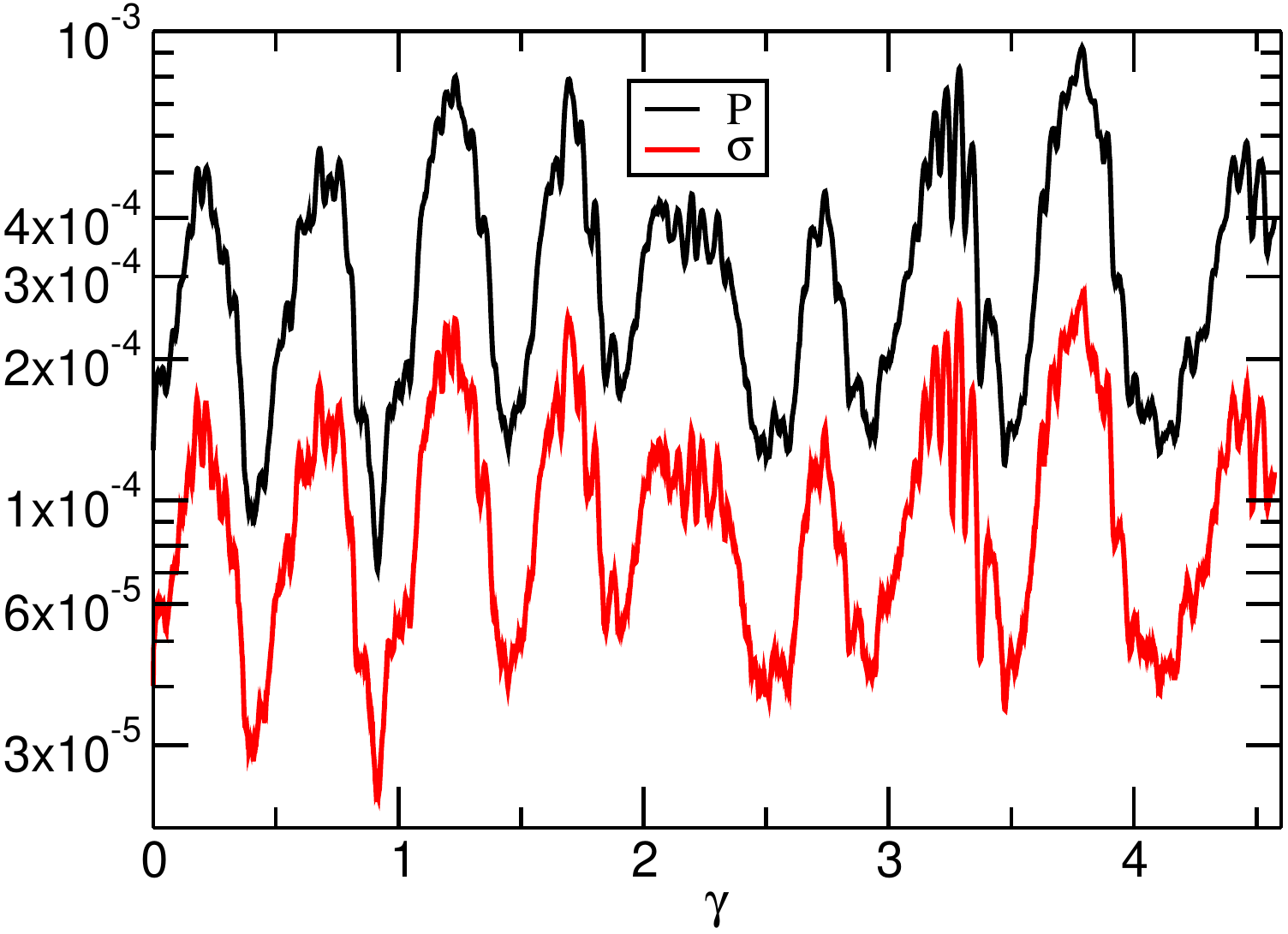}\vspace{2mm}
        \caption{Global pressure $p$ (black) and global stress
          $\sigma$ (red) as a function of strain $\gamma$.} \label{fig.pressure}
\end{figure}

The creation/destruction of the bands is also visible in the
time-dependent global stress and pressure signals, as shown in
Fig.~\ref{fig.pressure}. Both stress and pressure are highly
correlated, in fact $\sigma = \mu_{\rm eff} p$ with an effective
friction coefficient $\mu_{\rm eff}\approx 0.3$, quite similar to
other studies~\cite{peyneauPRE2008}.
Whenever a band is present the global stress is high. Apparently, the
signal displays some regularity, periodic oscillations with period of
$\Gamma\approx 0.5\gamma$, i.e. a strain of $50\%$. We have checked
that also the auto-correlation of the signal displays regular
oscillations. Simulations performed without periodic boundary
conditions in y-direction (with moving walls) also yield both
propagating clusters and bands. The regularity of the oscillations,
however, is strongly reduced.

\begin{figure}
  \includegraphics[width=0.51\textwidth]{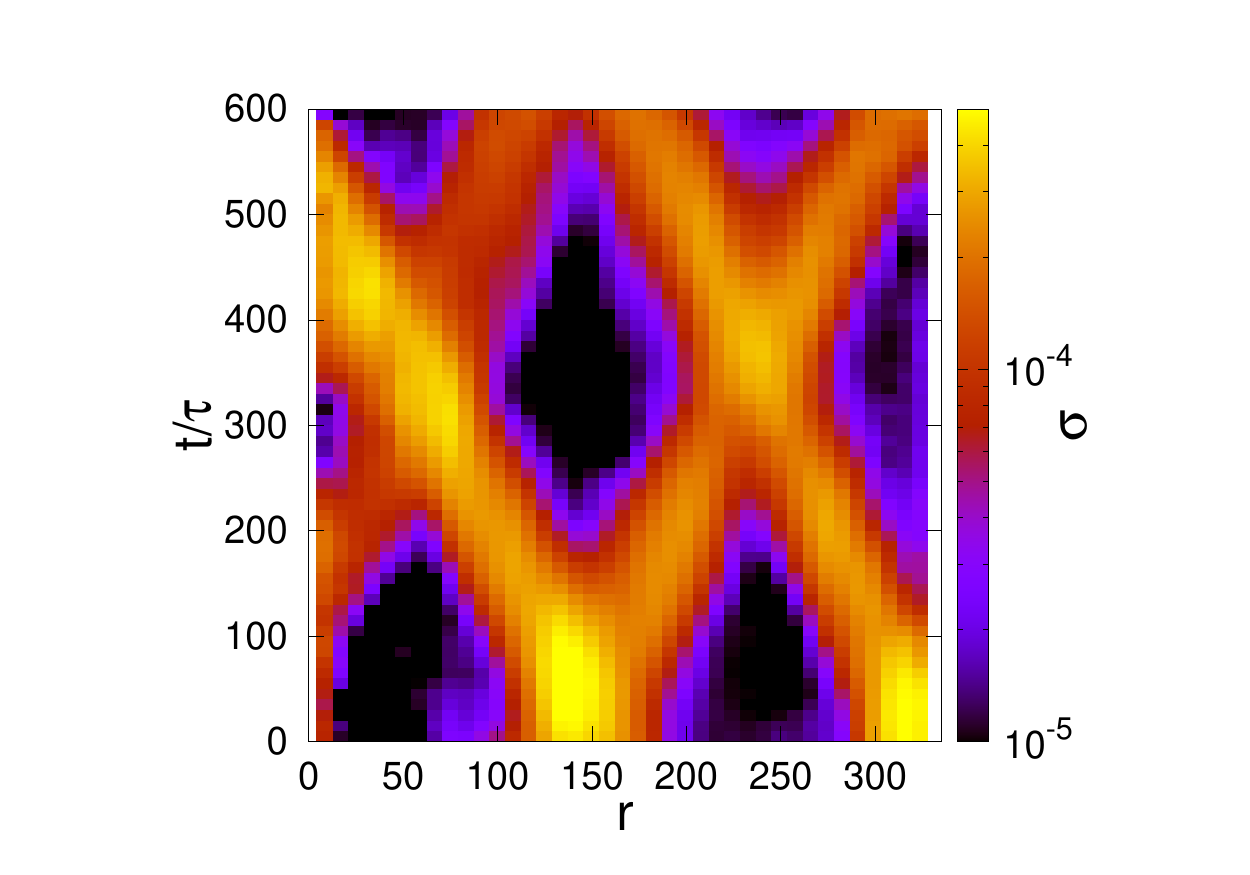}\\
  \includegraphics[width=0.51\textwidth]{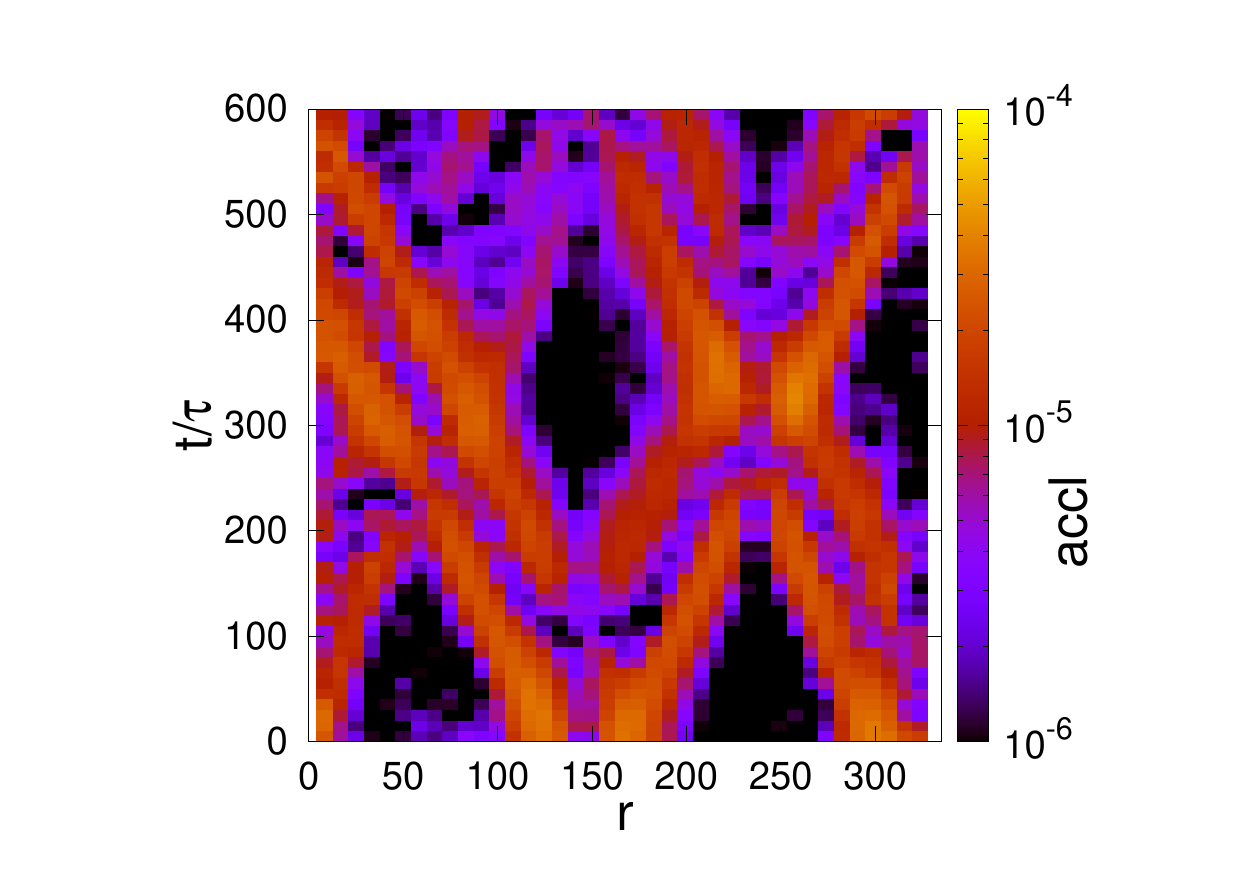}\\
 \includegraphics[width=0.51\textwidth]{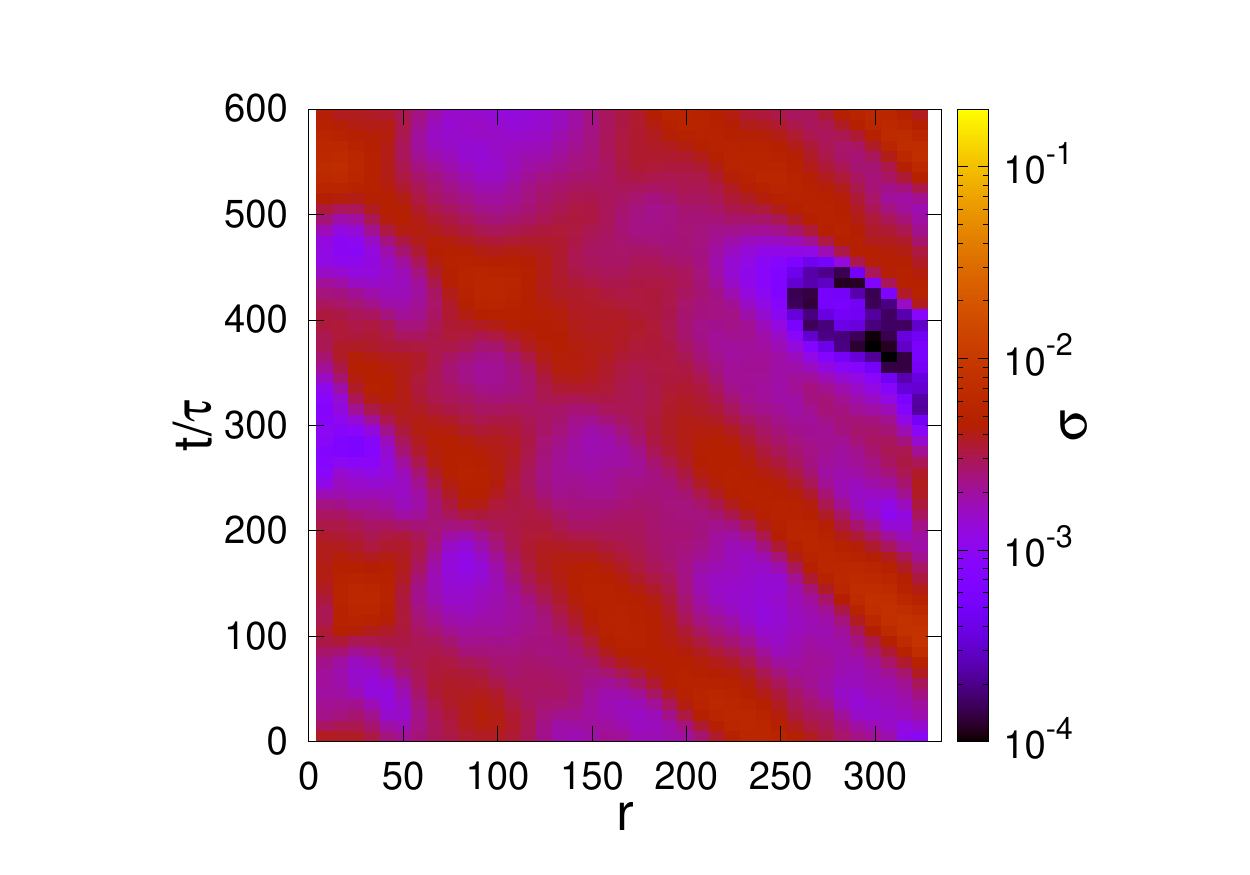}\\
  \caption{(upper panel) Time evolution of stress bands at the lower end of
    ST ($\dot\gamma=1.1\cdot 10^{-4}$). The x-axis represents the
    spatial coordinate along the dilational direction, i.e. along the
    propagation direction of the band; values are averaged over
    perpendicular stripes. The y-axis is time. Color code is shear
    stress. Four system-spanning bands move along dilational direction
    but in opposite directions, collide and move on. (middle panel) The same
    event, now with particle accelerations. (lower panel) Stress at the
    upper end of ST ($\dot\gamma=1.0\cdot 10^{-3}$). The x-axis now
    represents the spatial coordinate along the compressional
    direction; values are averaged over perpendicular stripes.}
  \label{fig:band_velocity}
\end{figure}

We can characterize the time evolution of the bands by plotting the
stress along the propagation (dilational) direction as a function of
time. Such a kymograph is shown in Fig.~\ref{fig:band_velocity} in the
first panel. Each horizontal layer gives the instantaneous stress
along the dilational direction and averaged over the compressional
direction. Moving upwards layer by layer increases time. Motion of a
band is therefore observed as a shift to the right or to the
left. Four bands can be seen, two of which move to the left, two move
to the right. In the X-shaped regions two bands collide and
subsequently separate again without delay. From such kymographs band
velocities can be estimated to be $|v^{band}|\approx 0.265 $, with
bands moving in both directions~\footnote{Band velocities of four
  bands are found to be   0.239, 0.269, -0.239 and -0.299. Negative
  sign indicates that band   moves away from the top layer.}.

We also display particle accelerations of the same sequence
(Fig.~\ref{fig:band_velocity} middle panel). It is clear that
accelerations are large at the edge of the bands, where the stress
gradient is high. This is a consequence of Newton's equation of
motion, or in continuum form, Navier-Stokes equation,
$\partial \vec v /\partial t \propto \nabla \sigma$.

In the right panel we display the propagation of the clusters visible
at larger strainrates $\dot\gamma=\dot\gamma_u$ (panel (c) of
Fig.~\ref{local_shear}). Now the propagation is along the
compressive direction, which is therefore taken as the spatial axis in
the plot. Velocities of these clusters are a factor of two to three
higher than for bands, $|v^{cl}|\approx 0.70$ ~\footnote{For three
  bands the velocities come out to be -0.67, -0.62 and -0.78.}. This
velocity is on the order of the sound velocity
$c_s = \sqrt{kd^2/m}=0.85$, with the relevant modulus taken as the
spring constant $k^{(n)}$.

Thus, we propose to categorize the dynamics as follows. In the Bagnold
regime at small strainrates, individual force-chains of a typical size
form. They are oriented along the compressive direction, and move
along this axis with the speed of sound. In the ST regime these
force-chains join together to form larger clusters that still move in
the compressive direction. The clusters themselves can join together,
percolating along the compressive direction. Further motion in this
direction is now blocked and the system could jam. However, associated
stress/pressure gradients at the edges of the bands induce a new
instability that leads to motion perpendicular to the band axis, in
direction of the gradient. On a local scale this motion may be
initiated via buckling-like events. Once started this motion is
sustained, because the stress gradient drives particle
accelerations. The resulting particle motion slightly densifies
surrounding region, and allows more contacts to form. These new
contacts are the foundation on which to build the large stress values,
which are encountered in the center of the band.

\section{Conclusion}\label{sec:conclusion}

A dense two-dimensional fluid of frictional grains undergoes a large
scale instability for strainrates which are intermediate between the
Bagnold and the HB regime. The unstable region is seen in $s-$shaped
flow curves in experiment~\cite{PhysRevE.92.032202} and simulations
and has been predicted within a simple hydrodynamic
analysis~\cite{GrobPRE2014}. In the unstable region of strain and
stress no stationary, homogenous flow is possible. What is observed
instead, depends crucially on system size. In the small system, we
observe hysteretic jumps between the two branches, the lower stress
value corresponding approximately to Bagnold flow and the larger one
to HB flow. Shear thickening is thus discontinuous. In a large system
both stress values are present simultaneously in large bands or
clusters. Such a state cannot be stationary, but is necessarily
time-dependent. At first sight, one might expect diffusive behaviour;
however, diffusion tends to equalize the stress profiles, implying
intermediate values of stress which are not allowed. Hence the bands
propagate in a wavelike fashion with a speed comparable to the speed
of sound. A spatial average over such heterogeneous stress states
gives rise to continuous shear thickening, so that in sufficiently
large and presumably infinite systems ST is always continuous.
What system size is necessary, depends on the
volume-fraction and therefore on the distance to the jamming
transition. Away from jamming, where only mild ST occurs, relatively
small systems suffice, while closer to jamming, larger and larger
systems are necessary to obtain a continuous flowcurve with a
continuous shear thickening regime.

By looking closer at the unstable regime, we find that the flow
is 
is qualitatively different at the upper and lower end of the unstable region.
At the upper end of the ST regime,
at large strainrates, we find large clusters, in which stress (and
pressure) is by orders of magnitude larger than outside of the
clusters. On the local scale these clusters consist of force-chains
oriented along the compressive direction of the flow. Clusters are not
stationary objects but move along the force-chain direction. They form
and dissolve frequently. 
At smaller strainrates, towards the lower end of the ST regime, we
find another large-scale structure: system-spanning stress
bands. These bands span the system along the compressive
direction. They seem to form out of percolating clusters, but their
motion is perpendicular to the force-chains, along the dilational
direction.

We conclude that large-scale instabilities give rise to continuous
flowcurves, if spatially averaged. The heterogeneous time-dependent
patterns come in with infinite correlation length, in the compressive
direction at the lower end and in the tensile direction at the upper
end.  This sudden onset is reminiscent of the appearance of
system-spanning plugs in impact experiments.

\begin{acknowledgments}
  We acknowledge financial support by the German Science Foundation
  (DFG) via the Heisenberg program (CH: HE-6322/2) and the SFB 937
  (SS); AZ acknowledges support by the Heraeus-Stiftung. 
\end{acknowledgments}
 

\begin{thebibliography}{32}%
\makeatletter
\providecommand \@ifxundefined [1]{%
 \@ifx{#1\undefined}
}%
\providecommand \@ifnum [1]{%
 \ifnum #1\expandafter \@firstoftwo
 \else \expandafter \@secondoftwo
 \fi
}%
\providecommand \@ifx [1]{%
 \ifx #1\expandafter \@firstoftwo
 \else \expandafter \@secondoftwo
 \fi
}%
\providecommand \natexlab [1]{#1}%
\providecommand \enquote  [1]{``#1''}%
\providecommand \bibnamefont  [1]{#1}%
\providecommand \bibfnamefont [1]{#1}%
\providecommand \citenamefont [1]{#1}%
\providecommand \href@noop [0]{\@secondoftwo}%
\providecommand \href [0]{\begingroup \@sanitize@url \@href}%
\providecommand \@href[1]{\@@startlink{#1}\@@href}%
\providecommand \@@href[1]{\endgroup#1\@@endlink}%
\providecommand \@sanitize@url [0]{\catcode `\\12\catcode `\$12\catcode
  `\&12\catcode `\#12\catcode `\^12\catcode `\_12\catcode `\%12\relax}%
\providecommand \@@startlink[1]{}%
\providecommand \@@endlink[0]{}%
\providecommand \url  [0]{\begingroup\@sanitize@url \@url }%
\providecommand \@url [1]{\endgroup\@href {#1}{\urlprefix }}%
\providecommand \urlprefix  [0]{URL }%
\providecommand \Eprint [0]{\href }%
\providecommand \doibase [0]{http://dx.doi.org/}%
\providecommand \selectlanguage [0]{\@gobble}%
\providecommand \bibinfo  [0]{\@secondoftwo}%
\providecommand \bibfield  [0]{\@secondoftwo}%
\providecommand \translation [1]{[#1]}%
\providecommand \BibitemOpen [0]{}%
\providecommand \bibitemStop [0]{}%
\providecommand \bibitemNoStop [0]{.\EOS\space}%
\providecommand \EOS [0]{\spacefactor3000\relax}%
\providecommand \BibitemShut  [1]{\csname bibitem#1\endcsname}%
\let\auto@bib@innerbib\@empty
\bibitem [{\citenamefont {Maharjan}\ \emph {et~al.}(2018)\citenamefont
  {Maharjan}, \citenamefont {Mukhopadhyay}, \citenamefont {Allen},
  \citenamefont {Storz},\ and\ \citenamefont {Brown}}]{maharjan18:_const}%
  \BibitemOpen
  \bibfield  {author} {\bibinfo {author} {\bibfnamefont {R.}~\bibnamefont
  {Maharjan}}, \bibinfo {author} {\bibfnamefont {S.}~\bibnamefont
  {Mukhopadhyay}}, \bibinfo {author} {\bibfnamefont {B.}~\bibnamefont {Allen}},
  \bibinfo {author} {\bibfnamefont {T.}~\bibnamefont {Storz}}, \ and\ \bibinfo
  {author} {\bibfnamefont {E.}~\bibnamefont {Brown}},\ }\href@noop {}
  {\bibfield  {journal} {\bibinfo  {journal} {Phys. Rev. E}\ }\textbf {\bibinfo
  {volume} {97}},\ \bibinfo {pages} {052602} (\bibinfo {year}
  {2018})}\BibitemShut {NoStop}%
\bibitem [{\citenamefont {Allen}\ \emph {et~al.}(2018)\citenamefont {Allen},
  \citenamefont {Sokol}, \citenamefont {Mukhopadhyay}, \citenamefont
  {Maharjan},\ and\ \citenamefont {Brown}}]{allen18:_system}%
  \BibitemOpen
  \bibfield  {author} {\bibinfo {author} {\bibfnamefont {B.}~\bibnamefont
  {Allen}}, \bibinfo {author} {\bibfnamefont {B.}~\bibnamefont {Sokol}},
  \bibinfo {author} {\bibfnamefont {S.}~\bibnamefont {Mukhopadhyay}}, \bibinfo
  {author} {\bibfnamefont {R.}~\bibnamefont {Maharjan}}, \ and\ \bibinfo
  {author} {\bibfnamefont {E.}~\bibnamefont {Brown}},\ }\href@noop {}
  {\bibfield  {journal} {\bibinfo  {journal} {Phys. Rev. E}\ }\textbf {\bibinfo
  {volume} {97}},\ \bibinfo {pages} {052603} (\bibinfo {year}
  {2018})}\BibitemShut {NoStop}%
\bibitem [{\citenamefont {Decker}\ \emph {et~al.}(2007)\citenamefont {Decker},
  \citenamefont {Halbach}, \citenamefont {Nam}, \citenamefont {Wagner},\ and\
  \citenamefont {Wetzel}}]{decker07:_stab_stf}%
  \BibitemOpen
  \bibfield  {author} {\bibinfo {author} {\bibfnamefont {M.~J.}\ \bibnamefont
  {Decker}}, \bibinfo {author} {\bibfnamefont {C.~J.}\ \bibnamefont {Halbach}},
  \bibinfo {author} {\bibfnamefont {C.~H.}\ \bibnamefont {Nam}}, \bibinfo
  {author} {\bibfnamefont {N.~J.}\ \bibnamefont {Wagner}}, \ and\ \bibinfo
  {author} {\bibfnamefont {E.~D.}\ \bibnamefont {Wetzel}},\ }\href@noop {}
  {\bibfield  {journal} {\bibinfo  {journal} {Composites Science and
  Technology}\ }\textbf {\bibinfo {volume} {67}},\ \bibinfo {pages} {565}
  (\bibinfo {year} {2007})}\BibitemShut {NoStop}%
\bibitem [{\citenamefont {Brown}\ and\ \citenamefont
  {Jaeger}(2012)}]{brown12JRheol}%
  \BibitemOpen
  \bibfield  {author} {\bibinfo {author} {\bibfnamefont {E.}~\bibnamefont
  {Brown}}\ and\ \bibinfo {author} {\bibfnamefont {H.}~\bibnamefont {Jaeger}},\
  }\href@noop {} {\bibfield  {journal} {\bibinfo  {journal} {J. Rheol.}\
  }\textbf {\bibinfo {volume} {56}},\ \bibinfo {pages} {875} (\bibinfo {year}
  {2012})}\BibitemShut {NoStop}%
\bibitem [{\citenamefont {Brown}\ and\ \citenamefont
  {Jaeger}()}]{0034-4885-77-4-046602}%
  \BibitemOpen
  \bibfield  {author} {\bibinfo {author} {\bibfnamefont {E.}~\bibnamefont
  {Brown}}\ and\ \bibinfo {author} {\bibfnamefont {H.~M.}\ \bibnamefont
  {Jaeger}},\ }\href@noop {} {\bibfield  {journal} {\bibinfo  {journal}
  {Reports on Progress in Physics}\ }\textbf {\bibinfo {volume} {77}},\
  \bibinfo {pages} {046602}}\BibitemShut {NoStop}%
\bibitem [{\citenamefont {Clavaud}\ \emph {et~al.}(2017)\citenamefont
  {Clavaud}, \citenamefont {B{\'e}rut}, \citenamefont {Metzger},\ and\
  \citenamefont {Forterre}}]{Clavaud5147}%
  \BibitemOpen
  \bibfield  {author} {\bibinfo {author} {\bibfnamefont {C.}~\bibnamefont
  {Clavaud}}, \bibinfo {author} {\bibfnamefont {A.}~\bibnamefont {B{\'e}rut}},
  \bibinfo {author} {\bibfnamefont {B.}~\bibnamefont {Metzger}}, \ and\
  \bibinfo {author} {\bibfnamefont {Y.}~\bibnamefont {Forterre}},\ }\href
  {\doibase 10.1073/pnas.1703926114} {\bibfield  {journal} {\bibinfo  {journal}
  {Proc. Natl. Acad. Sci. USA}\ }\textbf {\bibinfo {volume} {114}},\ \bibinfo
  {pages} {5147} (\bibinfo {year} {2017})},\ \Eprint
  {http://arxiv.org/abs/http://www.pnas.org/content/114/20/5147.full.pdf}
  {http://www.pnas.org/content/114/20/5147.full.pdf} \BibitemShut {NoStop}%
\bibitem [{\citenamefont {Comtet}\ \emph {et~al.}(2017)\citenamefont {Comtet},
  \citenamefont {Chatté}, \citenamefont {Niguès}, \citenamefont {Bocquet},
  \citenamefont {Siria},\ and\ \citenamefont {Colin}}]{comtet}%
  \BibitemOpen
  \bibfield  {author} {\bibinfo {author} {\bibfnamefont {J.}~\bibnamefont
  {Comtet}}, \bibinfo {author} {\bibfnamefont {G.}~\bibnamefont {Chatté}},
  \bibinfo {author} {\bibfnamefont {A.}~\bibnamefont {Niguès}}, \bibinfo
  {author} {\bibfnamefont {L.}~\bibnamefont {Bocquet}}, \bibinfo {author}
  {\bibfnamefont {A.}~\bibnamefont {Siria}}, \ and\ \bibinfo {author}
  {\bibfnamefont {A.}~\bibnamefont {Colin}},\ }\href
  {http://dx.doi.org/10.1038/ncomms15633} {\bibfield  {journal} {\bibinfo
  {journal} {Nat. Commun.}\ }\textbf {\bibinfo {volume} {8}},\ \bibinfo {pages}
  {15633} (\bibinfo {year} {2017})}\BibitemShut {NoStop}%
\bibitem [{\citenamefont {Pan}\ \emph {et~al.}(2015)\citenamefont {Pan},
  \citenamefont {de~Cagny}, \citenamefont {Weber},\ and\ \citenamefont
  {Bonn}}]{PhysRevE.92.032202}%
  \BibitemOpen
  \bibfield  {author} {\bibinfo {author} {\bibfnamefont {Z.}~\bibnamefont
  {Pan}}, \bibinfo {author} {\bibfnamefont {H.}~\bibnamefont {de~Cagny}},
  \bibinfo {author} {\bibfnamefont {B.}~\bibnamefont {Weber}}, \ and\ \bibinfo
  {author} {\bibfnamefont {D.}~\bibnamefont {Bonn}},\ }\href {\doibase
  10.1103/PhysRevE.92.032202} {\bibfield  {journal} {\bibinfo  {journal} {Phys.
  Rev. E}\ }\textbf {\bibinfo {volume} {92}},\ \bibinfo {pages} {032202}
  (\bibinfo {year} {2015})}\BibitemShut {NoStop}%
\bibitem [{\citenamefont {Hsu}\ \emph {et~al.}(2018)\citenamefont {Hsu},
  \citenamefont {Ramakrishna}, \citenamefont {Zanini}, \citenamefont
  {Spencer},\ and\ \citenamefont {Isa}}]{Hsu5117}%
  \BibitemOpen
  \bibfield  {author} {\bibinfo {author} {\bibfnamefont {C.-P.}\ \bibnamefont
  {Hsu}}, \bibinfo {author} {\bibfnamefont {S.~N.}\ \bibnamefont
  {Ramakrishna}}, \bibinfo {author} {\bibfnamefont {M.}~\bibnamefont {Zanini}},
  \bibinfo {author} {\bibfnamefont {N.~D.}\ \bibnamefont {Spencer}}, \ and\
  \bibinfo {author} {\bibfnamefont {L.}~\bibnamefont {Isa}},\ }\href {\doibase
  10.1073/pnas.1801066115} {\bibfield  {journal} {\bibinfo  {journal} {Proc.
  Natl. Acad. Sci. USA}\ }\textbf {\bibinfo {volume} {115}},\ \bibinfo {pages}
  {5117} (\bibinfo {year} {2018})},\ \Eprint
  {http://arxiv.org/abs/https://www.pnas.org/content/115/20/5117.full.pdf}
  {https://www.pnas.org/content/115/20/5117.full.pdf} \BibitemShut {NoStop}%
\bibitem [{\citenamefont {Heussinger}(2013)}]{ClausPRE2013}%
  \BibitemOpen
  \bibfield  {author} {\bibinfo {author} {\bibfnamefont {C.}~\bibnamefont
  {Heussinger}},\ }\href {\doibase 10.1103/PhysRevE.88.050201} {\bibfield
  {journal} {\bibinfo  {journal} {Phys. Rev. E}\ }\textbf {\bibinfo {volume}
  {88}},\ \bibinfo {pages} {050201} (\bibinfo {year} {2013})}\BibitemShut
  {NoStop}%
\bibitem [{\citenamefont {Seto}\ \emph {et~al.}(2013)\citenamefont {Seto},
  \citenamefont {Mari}, \citenamefont {Morris},\ and\ \citenamefont
  {Denn}}]{PhysRevLett.111.218301}%
  \BibitemOpen
  \bibfield  {author} {\bibinfo {author} {\bibfnamefont {R.}~\bibnamefont
  {Seto}}, \bibinfo {author} {\bibfnamefont {R.}~\bibnamefont {Mari}}, \bibinfo
  {author} {\bibfnamefont {J.~F.}\ \bibnamefont {Morris}}, \ and\ \bibinfo
  {author} {\bibfnamefont {M.~M.}\ \bibnamefont {Denn}},\ }\href {\doibase
  10.1103/PhysRevLett.111.218301} {\bibfield  {journal} {\bibinfo  {journal}
  {Phys. Rev. Lett.}\ }\textbf {\bibinfo {volume} {111}},\ \bibinfo {pages}
  {218301} (\bibinfo {year} {2013})}\BibitemShut {NoStop}%
\bibitem [{\citenamefont {Ness}\ and\ \citenamefont {Sun}(2016)}]{C5SM02326B}%
  \BibitemOpen
  \bibfield  {author} {\bibinfo {author} {\bibfnamefont {C.}~\bibnamefont
  {Ness}}\ and\ \bibinfo {author} {\bibfnamefont {J.}~\bibnamefont {Sun}},\
  }\href {\doibase 10.1039/C5SM02326B} {\bibfield  {journal} {\bibinfo
  {journal} {Soft Matter}\ }\textbf {\bibinfo {volume} {12}},\ \bibinfo {pages}
  {914} (\bibinfo {year} {2016})}\BibitemShut {NoStop}%
\bibitem [{\citenamefont {Wyart}\ and\ \citenamefont
  {Cates}(2014)}]{PhysRevLett.112.098302}%
  \BibitemOpen
  \bibfield  {author} {\bibinfo {author} {\bibfnamefont {M.}~\bibnamefont
  {Wyart}}\ and\ \bibinfo {author} {\bibfnamefont {M.~E.}\ \bibnamefont
  {Cates}},\ }\href {\doibase 10.1103/PhysRevLett.112.098302} {\bibfield
  {journal} {\bibinfo  {journal} {Phys. Rev. Lett.}\ }\textbf {\bibinfo
  {volume} {112}},\ \bibinfo {pages} {098302} (\bibinfo {year}
  {2014})}\BibitemShut {NoStop}%
\bibitem [{\citenamefont {Grob}\ \emph {et~al.}(2014)\citenamefont {Grob},
  \citenamefont {Heussinger},\ and\ \citenamefont {Zippelius}}]{GrobPRE2014}%
  \BibitemOpen
  \bibfield  {author} {\bibinfo {author} {\bibfnamefont {M.}~\bibnamefont
  {Grob}}, \bibinfo {author} {\bibfnamefont {C.}~\bibnamefont {Heussinger}}, \
  and\ \bibinfo {author} {\bibfnamefont {A.}~\bibnamefont {Zippelius}},\ }\href
  {\doibase 10.1103/PhysRevE.89.050201} {\bibfield  {journal} {\bibinfo
  {journal} {Phys. Rev. E}\ }\textbf {\bibinfo {volume} {89}},\ \bibinfo
  {pages} {050201} (\bibinfo {year} {2014})}\BibitemShut {NoStop}%
\bibitem [{\citenamefont {Fernandez}\ \emph {et~al.}(2013)\citenamefont
  {Fernandez}, \citenamefont {Mani}, \citenamefont {Rinaldi}, \citenamefont
  {Kadau}, \citenamefont {Mosquet}, \citenamefont {Lombois-Burger},
  \citenamefont {Cayer-Barrioz}, \citenamefont {Herrmann}, \citenamefont
  {Spencer},\ and\ \citenamefont {Isa}}]{PhysRevLett.111.108301}%
  \BibitemOpen
  \bibfield  {author} {\bibinfo {author} {\bibfnamefont {N.}~\bibnamefont
  {Fernandez}}, \bibinfo {author} {\bibfnamefont {R.}~\bibnamefont {Mani}},
  \bibinfo {author} {\bibfnamefont {D.}~\bibnamefont {Rinaldi}}, \bibinfo
  {author} {\bibfnamefont {D.}~\bibnamefont {Kadau}}, \bibinfo {author}
  {\bibfnamefont {M.}~\bibnamefont {Mosquet}}, \bibinfo {author} {\bibfnamefont
  {H.}~\bibnamefont {Lombois-Burger}}, \bibinfo {author} {\bibfnamefont
  {J.}~\bibnamefont {Cayer-Barrioz}}, \bibinfo {author} {\bibfnamefont {H.~J.}\
  \bibnamefont {Herrmann}}, \bibinfo {author} {\bibfnamefont {N.~D.}\
  \bibnamefont {Spencer}}, \ and\ \bibinfo {author} {\bibfnamefont
  {L.}~\bibnamefont {Isa}},\ }\href {\doibase 10.1103/PhysRevLett.111.108301}
  {\bibfield  {journal} {\bibinfo  {journal} {Phys. Rev. Lett.}\ }\textbf
  {\bibinfo {volume} {111}},\ \bibinfo {pages} {108301} (\bibinfo {year}
  {2013})}\BibitemShut {NoStop}%
\bibitem [{\citenamefont {Mari}\ \emph {et~al.}(2015)\citenamefont {Mari},
  \citenamefont {Seto}, \citenamefont {Morris},\ and\ \citenamefont
  {Denn}}]{Mari15326}%
  \BibitemOpen
  \bibfield  {author} {\bibinfo {author} {\bibfnamefont {R.}~\bibnamefont
  {Mari}}, \bibinfo {author} {\bibfnamefont {R.}~\bibnamefont {Seto}}, \bibinfo
  {author} {\bibfnamefont {J.~F.}\ \bibnamefont {Morris}}, \ and\ \bibinfo
  {author} {\bibfnamefont {M.~M.}\ \bibnamefont {Denn}},\ }\href {\doibase
  10.1073/pnas.1515477112} {\bibfield  {journal} {\bibinfo  {journal} {Proc.
  Natl. Acad. Sci. USA}\ }\textbf {\bibinfo {volume} {112}},\ \bibinfo {pages}
  {15326} (\bibinfo {year} {2015})},\ \Eprint
  {http://arxiv.org/abs/https://www.pnas.org/content/112/50/15326.full.pdf}
  {https://www.pnas.org/content/112/50/15326.full.pdf} \BibitemShut {NoStop}%
\bibitem [{\citenamefont {Grob}\ \emph {et~al.}(2016)\citenamefont {Grob},
  \citenamefont {Zippelius},\ and\ \citenamefont {Heussinger}}]{Grob2014}%
  \BibitemOpen
  \bibfield  {author} {\bibinfo {author} {\bibfnamefont {M.}~\bibnamefont
  {Grob}}, \bibinfo {author} {\bibfnamefont {A.}~\bibnamefont {Zippelius}}, \
  and\ \bibinfo {author} {\bibfnamefont {C.}~\bibnamefont {Heussinger}},\
  }\href {\doibase 10.1103/PhysRevE.93.030901} {\bibfield  {journal} {\bibinfo
  {journal} {Phys. Rev. E}\ }\textbf {\bibinfo {volume} {93}},\ \bibinfo
  {pages} {030901} (\bibinfo {year} {2016})}\BibitemShut {NoStop}%
\bibitem [{\citenamefont {Maiti}\ \emph {et~al.}(2016)\citenamefont {Maiti},
  \citenamefont {Zippelius},\ and\ \citenamefont
  {Heussinger}}]{0295-5075-115-5-54006}%
  \BibitemOpen
  \bibfield  {author} {\bibinfo {author} {\bibfnamefont {M.}~\bibnamefont
  {Maiti}}, \bibinfo {author} {\bibfnamefont {A.}~\bibnamefont {Zippelius}}, \
  and\ \bibinfo {author} {\bibfnamefont {C.}~\bibnamefont {Heussinger}},\
  }\href {http://stacks.iop.org/0295-5075/115/i=5/a=54006} {\bibfield
  {journal} {\bibinfo  {journal} {EPL (Europhysics Letters)}\ }\textbf
  {\bibinfo {volume} {115}},\ \bibinfo {pages} {54006} (\bibinfo {year}
  {2016})}\BibitemShut {NoStop}%
\bibitem [{\citenamefont {Chacko}\ \emph {et~al.}(2018)\citenamefont {Chacko},
  \citenamefont {Mari}, \citenamefont {Cates},\ and\ \citenamefont
  {Fielding}}]{PhysRevLett.121.108003}%
  \BibitemOpen
  \bibfield  {author} {\bibinfo {author} {\bibfnamefont {R.~N.}\ \bibnamefont
  {Chacko}}, \bibinfo {author} {\bibfnamefont {R.}~\bibnamefont {Mari}},
  \bibinfo {author} {\bibfnamefont {M.~E.}\ \bibnamefont {Cates}}, \ and\
  \bibinfo {author} {\bibfnamefont {S.~M.}\ \bibnamefont {Fielding}},\ }\href
  {\doibase 10.1103/PhysRevLett.121.108003} {\bibfield  {journal} {\bibinfo
  {journal} {Phys. Rev. Lett.}\ }\textbf {\bibinfo {volume} {121}},\ \bibinfo
  {pages} {108003} (\bibinfo {year} {2018})}\BibitemShut {NoStop}%
\bibitem [{\citenamefont {Saint-Michel}\ \emph {et~al.}(2018)\citenamefont
  {Saint-Michel}, \citenamefont {Gibaud},\ and\ \citenamefont
  {Manneville}}]{PhysRevX.8.031006}%
  \BibitemOpen
  \bibfield  {author} {\bibinfo {author} {\bibfnamefont {B.}~\bibnamefont
  {Saint-Michel}}, \bibinfo {author} {\bibfnamefont {T.}~\bibnamefont
  {Gibaud}}, \ and\ \bibinfo {author} {\bibfnamefont {S.}~\bibnamefont
  {Manneville}},\ }\href {\doibase 10.1103/PhysRevX.8.031006} {\bibfield
  {journal} {\bibinfo  {journal} {Phys. Rev. X}\ }\textbf {\bibinfo {volume}
  {8}},\ \bibinfo {pages} {031006} (\bibinfo {year} {2018})}\BibitemShut
  {NoStop}%
\bibitem [{\citenamefont {Rathee}\ \emph {et~al.}(2017)\citenamefont {Rathee},
  \citenamefont {Blair},\ and\ \citenamefont {Urbach}}]{rathee17:_local}%
  \BibitemOpen
  \bibfield  {author} {\bibinfo {author} {\bibfnamefont {V.}~\bibnamefont
  {Rathee}}, \bibinfo {author} {\bibfnamefont {D.~L.}\ \bibnamefont {Blair}}, \
  and\ \bibinfo {author} {\bibfnamefont {J.~S.}\ \bibnamefont {Urbach}},\
  }\href@noop {} {\bibfield  {journal} {\bibinfo  {journal} {Proc. Natl. Acad.
  Sci. (USA)}\ }\textbf {\bibinfo {volume} {114}},\ \bibinfo {pages} {8740}
  (\bibinfo {year} {2017})}\BibitemShut {NoStop}%
\bibitem [{\citenamefont {Silbert}\ \emph {et~al.}(2001)\citenamefont
  {Silbert}, \citenamefont {Ertas}, \citenamefont {Grest}, \citenamefont
  {Halsey}, \citenamefont {Levine},\ and\ \citenamefont {Plimpton}}]{silbert}%
  \BibitemOpen
  \bibfield  {author} {\bibinfo {author} {\bibfnamefont {L.~E.}\ \bibnamefont
  {Silbert}}, \bibinfo {author} {\bibfnamefont {D.}~\bibnamefont {Ertas}},
  \bibinfo {author} {\bibfnamefont {G.~S.}\ \bibnamefont {Grest}}, \bibinfo
  {author} {\bibfnamefont {T.~C.}\ \bibnamefont {Halsey}}, \bibinfo {author}
  {\bibfnamefont {D.}~\bibnamefont {Levine}}, \ and\ \bibinfo {author}
  {\bibfnamefont {S.~J.}\ \bibnamefont {Plimpton}},\ }\href {\doibase
  10.1103/PhysRevE.64.051302} {\bibfield  {journal} {\bibinfo  {journal} {Phys.
  Rev. E}\ }\textbf {\bibinfo {volume} {64}},\ \bibinfo {pages} {051302}
  (\bibinfo {year} {2001})}\BibitemShut {NoStop}%
\bibitem [{\citenamefont {Otsuki}\ and\ \citenamefont
  {Hayakawa}(2011)}]{Hayakawa}%
  \BibitemOpen
  \bibfield  {author} {\bibinfo {author} {\bibfnamefont {M.}~\bibnamefont
  {Otsuki}}\ and\ \bibinfo {author} {\bibfnamefont {H.}~\bibnamefont
  {Hayakawa}},\ }\href {\doibase 10.1103/PhysRevE.83.051301} {\bibfield
  {journal} {\bibinfo  {journal} {Phys. Rev. E}\ }\textbf {\bibinfo {volume}
  {83}},\ \bibinfo {pages} {051301} (\bibinfo {year} {2011})}\BibitemShut
  {NoStop}%
\bibitem [{\citenamefont {Plimpton}(1995)}]{lammps}%
  \BibitemOpen
  \bibfield  {author} {\bibinfo {author} {\bibfnamefont {S.}~\bibnamefont
  {Plimpton}},\ }\href {\doibase https://doi.org/10.1006/jcph.1995.1039}
  {\bibfield  {journal} {\bibinfo  {journal} {Journal of Computational
  Physics}\ }\textbf {\bibinfo {volume} {117}},\ \bibinfo {pages} {1 }
  (\bibinfo {year} {1995})},\ \bibinfo {note}
  {http://lammps.sandia.gov}\BibitemShut {NoStop}%
\bibitem [{\citenamefont {Irani}\ \emph {et~al.}(2018)\citenamefont {Irani},
  \citenamefont {Chaudhuri},\ and\ \citenamefont {Heussinger}}]{2018irani}%
  \BibitemOpen
  \bibfield  {author} {\bibinfo {author} {\bibfnamefont {E.}~\bibnamefont
  {Irani}}, \bibinfo {author} {\bibfnamefont {P.}~\bibnamefont {Chaudhuri}}, \
  and\ \bibinfo {author} {\bibfnamefont {C.}~\bibnamefont {Heussinger}},\
  }\href@noop {} {\bibfield  {journal} {\bibinfo  {journal} {ArXiv e-prints}\ }
  (\bibinfo {year} {2018})},\ \Eprint {http://arxiv.org/abs/1809.06128}
  {arXiv:1809.06128 [cond-mat.soft]} \BibitemShut {NoStop}%
\bibitem [{\citenamefont {Lois}\ \emph {et~al.}(2007)\citenamefont {Lois},
  \citenamefont {Lema\^{\i}tre},\ and\ \citenamefont
  {Carlson}}]{PhysRevE.76.021302}%
  \BibitemOpen
  \bibfield  {author} {\bibinfo {author} {\bibfnamefont {G.}~\bibnamefont
  {Lois}}, \bibinfo {author} {\bibfnamefont {A.}~\bibnamefont {Lema\^{\i}tre}},
  \ and\ \bibinfo {author} {\bibfnamefont {J.~M.}\ \bibnamefont {Carlson}},\
  }\href {\doibase 10.1103/PhysRevE.76.021302} {\bibfield  {journal} {\bibinfo
  {journal} {Phys. Rev. E}\ }\textbf {\bibinfo {volume} {76}},\ \bibinfo
  {pages} {021302} (\bibinfo {year} {2007})}\BibitemShut {NoStop}%
\bibitem [{\citenamefont {Cates}\ \emph {et~al.}(1999)\citenamefont {Cates},
  \citenamefont {Wittmer}, \citenamefont {Bouchaud},\ and\ \citenamefont
  {Claudin}}]{doi:10.1063/1.166456}%
  \BibitemOpen
  \bibfield  {author} {\bibinfo {author} {\bibfnamefont {M.~E.}\ \bibnamefont
  {Cates}}, \bibinfo {author} {\bibfnamefont {J.~P.}\ \bibnamefont {Wittmer}},
  \bibinfo {author} {\bibfnamefont {J.-P.}\ \bibnamefont {Bouchaud}}, \ and\
  \bibinfo {author} {\bibfnamefont {P.}~\bibnamefont {Claudin}},\ }\href
  {\doibase 10.1063/1.166456} {\bibfield  {journal} {\bibinfo  {journal}
  {Chaos: An Interdisciplinary Journal of Nonlinear Science}\ }\textbf
  {\bibinfo {volume} {9}},\ \bibinfo {pages} {511} (\bibinfo {year} {1999})},\
  \Eprint {http://arxiv.org/abs/https://doi.org/10.1063/1.166456}
  {https://doi.org/10.1063/1.166456} \BibitemShut {NoStop}%
\bibitem [{\citenamefont {Majmudar}\ and\ \citenamefont
  {Behringer}(2005)}]{majmudar05:_contac}%
  \BibitemOpen
  \bibfield  {author} {\bibinfo {author} {\bibfnamefont {T.~S.}\ \bibnamefont
  {Majmudar}}\ and\ \bibinfo {author} {\bibfnamefont {R.~P.}\ \bibnamefont
  {Behringer}},\ }\href@noop {} {\bibfield  {journal} {\bibinfo  {journal}
  {Nature}\ }\textbf {\bibinfo {volume} {435}},\ \bibinfo {pages} {1079}
  (\bibinfo {year} {2005})}\BibitemShut {NoStop}%
\bibitem [{\citenamefont {Guazzelli}\ and\ \citenamefont
  {Pouliquen}(2018)}]{guazzelli_pouliquen_2018}%
  \BibitemOpen
  \bibfield  {author} {\bibinfo {author} {\bibfnamefont {E.}~\bibnamefont
  {Guazzelli}}\ and\ \bibinfo {author} {\bibfnamefont {O.}~\bibnamefont
  {Pouliquen}},\ }\href {\doibase 10.1017/jfm.2018.548} {\bibfield  {journal}
  {\bibinfo  {journal} {Journal of Fluid Mechanics}\ }\textbf {\bibinfo
  {volume} {852}},\ \bibinfo {pages} {P1} (\bibinfo {year} {2018})}\BibitemShut
  {NoStop}%
\bibitem [{\citenamefont {Peyneau}\ and\ \citenamefont
  {Roux}(2008)}]{peyneauPRE2008}%
  \BibitemOpen
  \bibfield  {author} {\bibinfo {author} {\bibfnamefont {P.-E.}\ \bibnamefont
  {Peyneau}}\ and\ \bibinfo {author} {\bibfnamefont {J.-N.}\ \bibnamefont
  {Roux}},\ }\href@noop {} {\bibfield  {journal} {\bibinfo  {journal} {Phys.
  Rev. E}\ }\textbf {\bibinfo {volume} {78}},\ \bibinfo {pages} {011307}
  (\bibinfo {year} {2008})}\BibitemShut {NoStop}%
\bibitem [{Note1()}]{Note1}%
  \BibitemOpen
  \bibinfo {note} {Band velocities of four bands are found to be \IeC {\protect
  \nobreakspace } 0.239, 0.269, -0.239 and -0.299. Negative sign indicates that
  band \IeC {\protect \nobreakspace } moves away from the top
  layer.}\BibitemShut {Stop}%
\bibitem [{Note2()}]{Note2}%
  \BibitemOpen
  \bibinfo {note} {For three \IeC {\protect \nobreakspace } bands the
  velocities come out to be -0.67, -0.62 and -0.78.}\BibitemShut {Stop}%
\end{thebibliography}

%

\end{document}